\documentclass[aps,pra,reprint,superscriptaddress,longbibliography,nofootinbib]{revtex4-2}

\usepackage{amsmath,amssymb,amsfonts,bm,mathtools}
\usepackage{graphicx}
\usepackage{physics}
\usepackage{hyperref}
\usepackage{xcolor}
\usepackage{booktabs}
\usepackage{enumitem}

\hypersetup{
  colorlinks=true,
  linkcolor=blue,
  citecolor=blue,
  urlcolor=blue
}

\begin{document}

\title{Optimal Null-Constrained Source-Basis Sensing in a Time-Reversed Young Interferometer}

\author{Jianming Wen}
\email{jwen7@binghamton.edu}
\affiliation{Department of Electrical and Computer Engineering, Binghamton University, Binghamton, New York 13902, USA}

\date{\today}

\begin{abstract}
We develop a general theory of null-constrained parameter estimation in a time-reversed Young (TRY) interferometer, where measurement is performed through programmable source-basis encoding with a fixed detector. We address the fundamental question of how to design source patterns that enforce a true metrological null---vanishing nominal response at the operating point---while preserving finite first-order sensitivity to the parameter. Under a general shot-noise-limited channel model, we show that the optimal null-constrained receiver is obtained by projecting the derivative response onto the subspace orthogonal to the nominal background in the inverse-noise metric. This yields a constructive solution in which the optimal source-basis code is given by the inverse-noise-weighted derivative response with its background-parallel component removed. We further derive an exact and universal information-retention law: the locally accessible Fisher information is reduced by a factor $1-\chi^2$, where $\chi$ quantifies the inverse-noise overlap between the nominal and derivative response vectors. This result establishes a precise geometric interpretation of the cost of null enforcement. Numerical examples demonstrate the null-coded TRY receivers can retain nearly the full local information and can be accurately implemented using binary and positive-only source patterns. These findings identify source-basis null engineering as a distinct and practically viable modality for derivative-mode sensing, with implications for superresolution metrology and programmable optical measurement architectures.
\end{abstract}

\maketitle

\section{Introduction}
The fundamental limits of optical resolution are most fruitfully understood through parameter estimation rather than image formation alone. In this viewpoint, the central quantity is the Fisher information carried by the measurement record about the parameter of interest, which sets the ultimate precision through the Cram\'er--Rao bound \cite{Helstrom1976,Kay1993,VanTrees2004,Bettens1999,Ram2006,Tsang2016PRX}. This perspective has led to a broad reexamination of how optical measurements should be designed near symmetric points and other operating conditions where conventional intensity-based readout can become locally uninformative \cite{Tsang2016PRX,Nair2016PRL,Nair2016OE,Paur2016Optica,Tsang2019PRA}.

A recurring lesson from these developments is that the most useful local information is often carried not by the nominal intensity itself, but by how that intensity changes under a small parameter perturbation. Near a symmetry point, the direct signal can be dominated by a large background component, while the informative variation resides in derivative-like structure that is easily obscured in standard measurements \cite{Tsang2016PRX,Nair2016PRL,Nair2016OE,Paur2016Optica,Tsang2019PRA}. This observation underlies several important superresolution and precision-sensing strategies, including spatial-mode demultiplexing, parity-sensitive detection, and nulling-based receivers~\cite{Serabyn1999,SanchezSoto2017,Martinod2021NatComm,Wadood2021,Laugier2023}. Although these methods are physically diverse, they share a common principle: suppress the nominal background while preserving the component of the signal that changes most sensitively with the parameter.

In many such schemes, interference plays a decisive role because it determines whether the measurement record contains sufficiently sharp signed structure to support derivative-sensitive estimation. In particular, the contrast or visibility of the underlying interference pattern controls how strongly parameter-dependent redistribution appears across the available measurement channels. When interference visibility is weak, the response becomes dominated by a slowly varying positive background, and the derivative channel tends to align more strongly with the nominal signal. When visibility is stronger, the response develops more pronounced alternating structure, which can increase the derivative content that is distinguishable from the background and thereby improve the effectiveness of null-constrained sensing. Thus, visibility is not merely an experimental quality factor; it directly influences the geometry of the estimation problem.

A time-reversed Young (TRY) interferometer provides a fundamentally different physical realization of this general principle~\cite{Wen2025OC,Wen2026,Wen2026TRYdiff}. In contrast to the conventional Young geometry, where interference fringes are observed in the detection plane, the TRY configuration organizes the measurement directly in the source coordinate, using a fixed detector together with a programmable, point-addressable source plane~\cite{Wen2025OC,Wen2026,Wen2026TRYdiff}. This inversion turns the source plane itself into the measurement basis. As a result, measurement design is performed not through detector-side mode sorting or output-port engineering, but through controlled source-basis coding.

Recent work~\cite{Wen2026TRYdiff} showed that this source-basis programmability enables differential encoding, in which two positive source patterns are combined to synthesize an antisymmetric response channel and thereby realize derivative-sensitive local estimation. That result established that TRY can access informative first-order structure without raster-style scanning and can outperform simple source-plane intensity sampling. At the same time, it left open a more stringent and practically important question: how should the source basis be designed when the receiver is required to operate at a true metrological null?

This distinction is essential. A dark output alone does not guarantee a useful sensing configuration~\cite{Serabyn1999,Martinod2021NatComm,Laugier2023}. A single nonnegative intensity channel can vanish at an operating point yet remain even in the perturbation, so that its leading response is quadratic and does not preserve the sign of the parameter shift. By contrast, a metrologically useful null must satisfy two conditions simultaneously: the nominal response must vanish at the operating point, and the first derivative with respect to the parameter must remain finite. Such a receiver is not merely dark; it is locally informative.

In the TRY architecture, this requirement becomes a source-basis design problem. Because the primitive record is distributed across addressable source channels, the null need not be tied to any individual dark physical port. Instead, it can be synthesized by combining multiple finite-flux channels so that the nominal output cancels while the derivative response survives. This turns null operation into a constrained coding problem in response space: one seeks source-basis weights that reject the background but preserves as much first-order sensitivity as possible.

In this work, we develop a general theory of \emph{null-constrained source-basis sensing} in a TRY interferometer. Under a shot-noise-limited channel model, we show that the optimal null-coded receiver is obtained by projecting the derivative response onto the subspace orthogonal to the nominal background in the inverse-noise metric. Equivalently, the optimal code is the \emph{inverse-noise-weighted derivative response with its background-parallel component removed}. This identifies null-coded TRY metrology as a \emph{source-basis realization of derivative-mode sensing}~\cite{Nair2016OE,SanchezSoto2017,Tsang2019PRA}, implemented not through detection-mode projections but through programmable source-coordinate coding.

The developed theory also yields an exact and compact information-retention law. We show that the locally accessible Fisher information under the null constraint is reduced from the unconstrained full-channel value by the factor $1-\chi^2$, where $\chi$ is the inverse-noise-metric overlap between the nominal and derivative response vectors. This result gives a precise geometric characterization of when null operation is essentially lossless and when it incurs a substantial penalty.

Within this framework, interference visibility acquires a second and more structured meaning. Beyond controlling fringe contrast in the primitive channel pattern, it influences the overlap between the nominal response and the derivative response and therefore helps determine the information cost of null enforcement. Higher visibility generally sharpens the derivative-bearing interference structure and can reduce the degree to which the informative variation is masked by the nominal background, whereas lower visibility tends to wash out the signed interference content and can make the null-constrained projection less favorable. For this reason, visibility should be regarded as an estimation-relevant control parameter of the TRY architecture, not merely as a descriptive property of the underlying fringe pattern.

The remainder of the paper develops this framework systematically. Section II introduces the primitive source-basis channel model and clarifies the role of interference structure and visibility in shaping the local response manifold. Section III defines a true differential null and distinguishes it from one-port darkness. Section IV derives the optimal null code and the exact information-retention law. Section V presents explicit TRY examples and quantitative comparisons, including how interference visibility affects code structure, retained information, and robustness under detuning. Section VI analyzes binary and positive-only implementations, and Section VII places the results in the broader context of nulling, superresolution, and programmable measurement design.

\section{Primitive Source-Basis Channel Model}
A TRY interferometer is naturally described in a source-basis representation, in which a fixed detector is paired with a programmable, point-addressable source plane~\cite{Wen2025OC,Wen2026}. In this formulation, the measurement record is organized over source coordinate rather than an observation plane, and the source plane itself serves as the basis in which the measurement is defined. This representation makes clear that measurement design is carried out through source-basis coding rather than detector-side mode selection.

\subsection{General channel model}
We discretize the source plane into $M$ addressable channels centered at coordinates $\{x_m\}_{m=1}^M$. For a given value of the unknown scalar parameter $\theta$, the primitive measurement record consists of independent photon counts
\begin{equation}
n_m\sim\mathrm{Pois}(\lambda_m(\theta)), \quad m=1,\dots,M,
\end{equation}
where $\lambda_m(\theta)$ denotes the mean response of the $m$th source channel. Collecting these channel means into a vector,
\begin{equation}
\boldsymbol{\lambda}(\theta) =
\begin{pmatrix}
\lambda_1(\theta) \\
\lambda_2(\theta) \\
\vdots \\
\lambda_M(\theta)
\end{pmatrix},
\end{equation}
we obtain a complete statistical description of the primitive TRY measurement in the source basis~\cite{Wen2026TRYdiff}.

This formulation is deliberately model-independent. The null-coding theory developed below depends only on the local behavior of $\boldsymbol{\lambda}(\theta)$ in a neighborhood of a chosen operating point $\theta_0$. It therefore applies equally to analytically derived forward models, experimentally calibrated response functions, or numerically generated channel families.

\subsection{Local expansion and noise structure}
We define the nominal response and its first two derivatives at the operating point $\theta_0$ as
\begin{equation}
\boldsymbol{\lambda}_0=\boldsymbol{\lambda}(\theta_0), \quad
\boldsymbol{\lambda}_1=\left.\frac{d\boldsymbol{\lambda}}{d\theta}\right|_{\theta_0}, \quad
\boldsymbol{\lambda}_2=\left.\frac{d^2\boldsymbol{\lambda}}{d\theta^2}\right|_{\theta_0}.
\end{equation}
For a small perturbation $\delta=\theta-\theta_0$, the channel response admits the expansion
\begin{equation}
\boldsymbol{\lambda}(\theta_0+\delta)=\boldsymbol{\lambda}_0+\boldsymbol{\lambda}_1\delta+\frac{1}{2}\boldsymbol{\lambda}_2\delta^2+O(\delta^3).
\end{equation}
Under the independent Poisson model, the noise covariance is diagonal,
\begin{equation}
\mathbf{D}(\theta)=\mathrm{diag}\big(\lambda_m(\theta)\big)^M_{m=1}=
\mathrm{diag}\big(\lambda_1(\theta),\dots,\lambda_M(\theta)\big),
\end{equation}
and at the operating point $\theta_0$,
\begin{equation}
\mathbf{D}_0=\mathrm{diag}(\lambda_{0,1},\dots,\lambda_{0,M}).
\end{equation}
The Fisher information carried by the full primitive channel record at $\theta_0$ is thus
\begin{equation}
\mathcal{I}_{\mathrm{full}}(\theta_0)=\sum_{m=1}^M\frac{[\lambda_m'(\theta_0)]^2}{\lambda_m(\theta_0)}=\boldsymbol{\lambda}_1^{T}\mathbf{D}_0^{-1}\boldsymbol{\lambda}_1,
\end{equation}
whose quantity provides the natural local benchmark against which all coded receivers will be compared.

\subsection{Role of interference visibility in the local response geometry}
To make explicit how interference affects the null-coding problem, it is useful to introduce the phenomenological TRY forward law already used in earlier work~\cite{Wen2026TRYdiff},
\begin{equation}
\lambda(x;\theta)=N_0\eta\,g(x-\theta)\left[1+V\cos(\kappa x+\phi_0)\right],
\end{equation}
with a Gaussian localized envelope
\begin{equation}
g(x-\theta)=\exp\!\left[-\frac{(x-\theta)^2}{2\sigma^2}\right].
\end{equation}
Here $N_0$ is the mean launched photon number per source-addressing interval, $\eta$ is the overall detection efficiency, $V$ is the fringe visibility, $\kappa$ is the spatial frequency of the source-plane interference, and $\phi_0$ is a phase offset, and $\sigma$ is the width of the localized source-plane response. Moreover, for simplicity, $\theta$ here parameterizes the displacement of the localized source-response envelope relative to a fixed source-basis interference modulation; accordingly, the fringe factor $\cos(kx+\phi_0)$ is taken to be stationary in the source coordinate. For the case where $\theta$ is meant to represent a literal translation of the entire source-plane intensity pattern, we leave the detailed treatment as an exercise to the reader. 

Sampling this field at the channel centers at $x_m$ gives the primitive means $\lambda_m(\theta)=\lambda(x_m;\theta)$. At the operating point $\theta_0$,
\begin{equation}
\lambda_{0,m}=N_0\eta\,g^{(0)}_m\left[1+V\cos(\kappa x_m+\phi_0)\right],
\end{equation}
where $g^{(0)}_m\equiv g(x_m-\theta_0)$. Because the parameter enters through a displacement of the localized envelope, the first derivative is
\begin{align}
\lambda_{1,m}&=\frac{\partial\lambda(x_m;\theta)}{\partial\theta}\bigg|_{\theta_0}\nonumber\\
&=N_0\eta\frac{x_m-\theta_0}{\sigma^2}g^{(0)}_m\left[1+V\cos(\kappa x_m+\phi_0)\right].
\end{align}
This expression already shows an important point: in the present model, the parameter derivative is multiplicatively modulated by the same interference factor that appears in the nominal response. The visibility thus enters not only through the signal amplitude, but through the detailed channel-by-channel structure of both $\boldsymbol{\lambda}_0$ and $\boldsymbol{\lambda}_1$.

The full-channel Fisher information (7) becomes
\begin{align}
\mathcal I_{\rm full}(\theta_0)&=\sum^M_{m=1}\frac{\lambda^2_{1,m}}{\lambda_{0,m}}\nonumber\\
&=N_0\eta\sum^M_{m=1}\frac{(x_m-\theta_0)^2}{\sigma^4}g^{(0)}_m\left[1+V\cos(\kappa x_m+\phi_0)\right].
\end{align}
Hence, within this forward model, visibility changes the local information content directly by reweighting the contribution of each source channel.

More importantly for the null-constrained problem developed later, visibility also affects the inverse-noise overlap between the nominal and derivative response vectors. The relevant inner product is
\begin{equation}
\boldsymbol{\lambda}^T_0\mathbf{D}^{-1}_0\boldsymbol{\lambda}_1=\sum^M_{m=1}\lambda_{1,m},
\end{equation}
which here becomes
\begin{equation}
\boldsymbol{\lambda}^T_0\mathbf{D}^{-1}_0\boldsymbol{\lambda}_1=N_0\eta\sum^M_{m=1}\frac{x_m-\theta_0}{\sigma^2}g^{(0)}_m\left[1+V\cos(\kappa x_m+\phi_0)\right].
\end{equation}
This term is precisely the background-aligned component of the derivative response that must be removed by the optimal null code. The visibility $V$ therefore influences the null penalty not abstractly, but through the weighted signed sum above.

Several consequences immediately follow:

(I) When $V$=0, the response reduces to a purely positive displaced envelope,
\begin{equation}
\lambda_{0,m}=N_0\eta\,g^{(0)}_m,\quad\lambda_{1,m}=N_0\eta\frac{x_m-\theta_0}{\sigma^2}g^{(0)}_m.
\end{equation}
In this limit, the local geometry is controlled entirely by the envelope translation. For symmetric sampling about $\theta_0$, the overlap sum can vanish by parity, but the derivative structure remains smooth and lacks interference-enhanced channel alternation.

(II) When $V>0$, the interference term redistributes both nominal flux and derivative weight across the source basis. This reshaping can either reduce or increase the overlap between $\boldsymbol{\lambda}_0$ and $\boldsymbol{\lambda}_1$, depending on $\kappa,\phi_0,\theta_0$, and the discretization. In particular, higher visibility tends to strengthen channel-to-channel contrast and can create derivative structure that is less background-like in the inverse-noise metric, thereby making null enforcement less costly.

(III) The quantity that matters later is not visibility alone, but how visibility modifies the normalized overlap
\begin{equation}
\chi=\frac{\boldsymbol{\lambda}_0^T\mathbf{D}_0^{-1}\boldsymbol{\lambda}_1}{\sqrt{(\boldsymbol{\lambda}_0^T\mathbf{D}_0^{-1}\boldsymbol{\lambda}_0)(\boldsymbol{\lambda}_1^T\mathbf{D}_0^{-1}\boldsymbol{\lambda}_1)}}.
\end{equation}
Accordingly, visibility should be viewed as a physical control parameter that shapes the response-space angle between $\boldsymbol{\lambda}_0$ and $\boldsymbol{\lambda}_1$, and thence the fraction of information retained under the null constraint. This quantitative link will be exploited later when we analyze the optimal null code and the exact retention law.

\subsection{Scope of the TRY forward model}

The explicit model above will be used in later sections only as an illustrative channel family for numerical examples and physical interpretation. None of the general derivations below depends on the Gaussian envelope or cosine modulation specifically. The analytical theory requires only the local quantities $\boldsymbol{\lambda}_0,\boldsymbol{\lambda}_1$, and $\mathbf{D}_0$. For this particular discretized model, however, the interference visibility also permits an exact analytical expression for the overlap parameter $\chi$, which is derived in Appendix~A and used later to interpret the visibility dependence of the numerical examples in Sec.~V.

\section{Definition of a True Metrological Null}

A source-basis receiver in the TRY architecture is specified by a real-valued code vector 
\begin{equation}
\mathbf{w}=(w_1,\dots,w_M)^T
\end{equation}
acting linearly on the primitive channel counts. The resulting coded observable is
\begin{equation}
S_{\mathbf{w}}=\sum_{m=1}^M w_mn_m=\mathbf{w}^T\mathbf{n},
\end{equation}
with mean and variance
\begin{equation}
\langle S_{\mathbf{w}}(\theta)\rangle=\mathbf{w}^T\boldsymbol{\lambda}(\theta), 
\quad \mathrm{Var}[S_{\mathbf{w}}(\theta)]=\mathbf{w}^T\mathbf{D}(\theta)\mathbf{w},
\end{equation}
where $\mathbf{D}(\theta)$ is given by Eq.~(5). This linear construction is the basic mechanism by which the TRY architecture turns programmable source patterns into effective measurement channels.

\subsection{Null condition and local informativeness}

Let $\theta_0$ denote the operating point. A \emph{true metrological null} should satisfy not only the vanishing nominal output, but also nonvanishing first-order sensitivity to the parameter. We hence define a null-constrained receiver by the pair of conditions
\begin{equation}
\mathbf{w}^T\boldsymbol{\lambda}_0=0, \quad 
\mathbf{w}^T\boldsymbol{\lambda}_1\neq 0,
\end{equation}
where, as in Sec. IIB, $\boldsymbol{\lambda}_0=\boldsymbol{\lambda}(\theta_0)$ and $\boldsymbol{\lambda}_1=\boldsymbol{\lambda}'(\theta_0)$. The first condition enforces exact cancellation of the nominal response at $\theta_0$. The second ensures that the coded observable remains locally informative: a small parameter displacement still produces a signed linear response.

Indeed, for $\delta=\theta-\theta_0$, the expansion of Sec. II gives
\begin{align}
\langle S_{\mathbf{w}}(\theta_0+\delta)\rangle&= \mathbf{w}^T\boldsymbol{\lambda}_0+\mathbf{w}^T\boldsymbol{\lambda}_1\,\delta
+\frac{1}{2}\mathbf{w}^T\boldsymbol{\lambda}_2\,\delta^2+O(\delta^3)\nonumber\\
&=\mathbf{w}^T\boldsymbol{\lambda}_1\,\delta
+\frac{1}{2}\mathbf{w}^T\boldsymbol{\lambda}_2\,\delta^2+O(\delta^3)\nonumber\\
&\approx\mathbf{w}^T\boldsymbol{\lambda}_1\,\delta+O(\delta^2).
\end{align}
Thus, a metrologically useful null is not merely dark: it is dark while preserving a signed linear response to small parameter displacements. This is the essential distinction between background suppression and local estimation capability.

This definition is stronger than merely requiring a small output. It identifies the operating point as one of deliberate background rejection together with retained first-order information. In the present source-basis setting, the null is therefore not a passive feature of the channel response, but an actively engineered property of the coded observable.

\subsection{Distinction from one-port darkness}
It is important to distinguish the present null condition from the more common notion of operating at a ``dark'' physical output~\cite{Serabyn1999,Martinod2021NatComm,Laugier2023}. A single nonnegative measurement channel $\mu(\theta)\ge0$ may vanish at $\theta_0$ and still fail to provide a useful signed first-order estimator. In a typical dark-port situation one has
\begin{equation}
\mu(\theta_0+\delta)=a\,\delta^2+O(\delta^4),
\end{equation}
so that the leading response is even in $\delta$.

Such a signal may still carry Fisher information in an idealized Poisson description, but it does not distinguish $\delta$ from $-\delta$ at leading order and therefore does not act as a linear local readout. By contrast, the null defined above is imposed on a differential observable synthesized from multiple finite-flux source channels, so that the nominal response cancels while the first-order derivative contribution survives. Thus, the relevant issue is not darkness alone, but whether the coded observable remains linearly informative about the parameter displacement. (The distinction between one-port darkness and the true differential null used here is discussed further in Appendix~B.)

This distinction is particularly important in the TRY setting. The objective is not merely to identify a low-output channel, but to construct a source-basis observable that rejects the background component of the record while preserving the part that changes linearly with the parameter.

\subsection{Source-basis interpretation and connection to response geometry}

Because the TRY architecture is programmable in the source coordinate, the null is not tied to a fixed interferometric port. Instead, it is engineered through the code vector $\bf w$. In response-space language, the condition $\mathbf{w}^T\boldsymbol{\lambda}_0=0$ means that the receiver is blind to the nominal source-basis response at $\theta_0$, whereas $\mathbf{w}^T\boldsymbol{\lambda}_1\neq0$ requires the same receiver to remain sensitive to the derivative direction.

This directly connects Sec.~III to the quantitative discussion in Sec.~II.C. There, the interference structure and visibility were shown to shape the local geometry of $\boldsymbol{\lambda}_0$ and $\boldsymbol{\lambda}_1$, and in particular the normalized overlap $\chi$ (see Eq.~(16)). A useful null is therefore possible only to the extent that the derivative response contains a component distinguishable from the nominal background in this inverse-noise geometry.

When the derivative direction is strongly aligned with the background, null enforcement necessarily removes a large portion of the useful signal. When the two are nearly orthogonal, the null can be imposed with much less penalty. In this sense, null operation is not a binary property but a constrained design problem whose quality is governed by the response-space angle between $\boldsymbol{\lambda}_0$ and $\boldsymbol{\lambda}_1$.

\subsection{Transition to optimal design}
The next question is therefore not whether a null can be imposed, but how to choose $\mathbf{w}$ so that the null-constrained receiver retains the maximum local sensitivity allowed by the primitive channel statistics. This requires optimizing the first-order response relative to the shot-noise level at the operating point. Section IV solves this problem exactly and shows that the optimal code is obtained by projecting the derivative response away from the nominal background in the inverse-noise metric.

\section{Optimal Null Code and Exact Information Retention}
\subsection{Local Fisher information of a coded observable}
We now determine the optimal source-basis code under the null constraint. For a small displacement $\delta=\theta-\theta_0$, in Sec.~III.A the coded observable responds as
\begin{equation*}
\langle S_{\bf w}(\theta_0+\delta)\rangle=\mathbf{w}^T\boldsymbol{\lambda}_1\delta+O(\delta^2),\;\mathrm{Var}[S_{\bf w}(\theta_0)]=\mathbf{w}^T\mathbf{D}_0\mathbf{w}.
\end{equation*}
In the local Gaussian small-signal approximation, the Fisher information carried by the scalar observable $S_{\bf w}$ at $\theta_0$ becomes
\begin{equation}
\mathcal I_{\bf w}(\theta_0)=\frac{[\partial_{\delta}\langle S_{\bf w}(\theta_0+\delta)\rangle]^2}{\mathrm{Var}[S_{\bf w}(\theta_0)]}=\frac{(\mathbf{w}^T\boldsymbol{\lambda}_1)^2}{\mathbf{w}^T\mathbf{D}_0\mathbf{w}}.
\end{equation}
This quantity should be distinguished from the full-channel Fisher information
\begin{equation}
\mathcal I_{\rm full}(\theta_0)=\boldsymbol{\lambda}^T_1\mathbf{D}^{-1}_0\boldsymbol{\lambda}_1,
\end{equation}
introduced in Sec.~II.B. The former quantifies the information captured by a specific coded receiver, whereas the latter is the total information available in the full primitive Poisson channel record.

The Equations presented above may also be motivated from the local signal-to-noise ratio,
\begin{equation}
\mathrm{SNR}_{\bf w}(\delta)\approx\frac{|\langle S_{\bf w}(\theta_0+\delta)\rangle|}{\sqrt{\mathrm{Var}[S_{\bf w}(\theta_0)]}}=|\delta|\frac{|\mathrm{w}^T\boldsymbol{\lambda}_1|}{\sqrt{\mathbf{w}^T\mathbf{D}_0\mathbf{w}}},
\end{equation}
so that $\mathcal I_{\bf w}(\theta_0)$ is precisely the squared slope-to-noise ratio governing local resolution.

Under the null constraint (20), the design problem becomes
\[
\max_{\bf w}\mathcal I_{\bf w}(\theta_0)\;\;\text{subject to}\;\;\mathbf{w}^T\boldsymbol{\lambda}_0=0.
\]
Because $\mathcal I_{\bf w}$ is invariant under an overall rescaling of $\bf w$, one may impose the normalization
\begin{equation}
\mathbf{w}^T\mathbf{D}_0\mathbf{w}=1
\end{equation}
without loss of generality. The optimization then becomes a constrained maximization of $\mathbf{w}^T\boldsymbol{\lambda}_1$. The detailed Lagrange-multiplier derivation can be found in Appendix~C; here we quote the result and then interpret it.

\subsection{Optimal null code and the role of $\alpha$}
The optimal null-constrained code is
\begin{equation}
\mathbf{w}^{\star}\propto\mathbf{D}^{-1}_0(\boldsymbol{\lambda}_1-\alpha\boldsymbol{\lambda}_0),
\end{equation}
where
\begin{equation}
\alpha=\frac{\boldsymbol{\lambda}^T_0\mathbf{D}^{-1}_0\boldsymbol{\lambda}_1}{\boldsymbol{\lambda}^T_0\mathbf{D}^{-1}_0\boldsymbol{\lambda}_0}.
\end{equation}
This is the central constructive result. The vector $\bf w^{\star}$ is the actual source-basis code used to form the optimal coded observable $S_{\bf w^{\star}}$; in Sec.~V it will be used directly to generate the source-basis patterns and sensitivity curves, including the code profiles shown in Fig.~3. So, $\bf w^{\star}$ is not just an intermediate algebraic object: it is the optimal receiver prescription produced by the theory.

The coefficient $\alpha$ has a clear geometric meaning. It is the inverse-noise-metric projection coefficient of the derivative response onto the nominal background. Equivalently, $\alpha\boldsymbol{\lambda}_0$ is exactly the background-aligned component of $\boldsymbol{\lambda}_1$ that must be removed in order to enforce the null. The larger $|\alpha|$ is, the larger the derivative component that is contaminated by the nominal response, and therefore the larger the penalty associated with null enforcement.

This also clarifies the connection to Sec.~II.C: the same overlap term $\boldsymbol{\lambda}^T_0\mathbf{D}^{-1}_0\boldsymbol{\lambda}_1$ in $\chi$ (Eq.~(16)) that entered the visibility-sensitive discussion there now appears directly in the optimal subtraction coefficient $\alpha$. Thus, interference structure and visibility influence the optimal null code through the amount of derivative response aligned with the background.

\subsection{Projection interpretation and relation to the full-channel limit}
In the absence of the null constraint, the optimal linear receiver aligns with the inverse-noise-weighted derivative direction $\mathbf{D}^{-1}_0\boldsymbol{\lambda}_1$, and the corresponding maximum information is
\begin{equation}
\mathcal I^{\rm lin}_{\rm max}=\boldsymbol{\lambda}^T_1\mathbf{D}^{-1}_0
\boldsymbol{\lambda}_1=\mathcal I_{\rm full}(\theta_0).
\end{equation}
The null constraint modifies this by requiring orthogonality to $\boldsymbol{\lambda}_0$. The optimal code is thus obtained by removing from $\mathbf{D}^{-1}_0\boldsymbol{\lambda}_1$ the component parallel to the nominal background. In this sense,
\begin{equation}
\mathbf{D}^{-1}_0\boldsymbol{\lambda}_1\to\mathbf{D}^{-1}_0(\boldsymbol{\lambda}_1-\alpha\boldsymbol{\lambda}_0)\propto\mathbf{w}^{\star}
\end{equation}
is an inverse-noise-weighted projection of the derivative response into the null-admissible subspace.

This projection picture is shown schematically in Fig.~1. The derivative response is decomposed into background-parallel and background-orthogonal parts in the inverse-noise metric; the latter is exactly the component retained under the null constraint. This geometric decomposition becomes fully explicit in the whitened representation introduced next.

\begin{figure}[hbt]
\includegraphics[width=0.55\columnwidth]{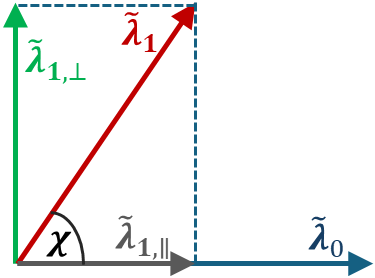}
\caption{Geometric interpretation of optimal null coding in inverse-noise-weighted response space. The derivative response $\tilde{\boldsymbol{\lambda}}_1$ is decomposed into a component parallel to the nominal background $\tilde{\boldsymbol{\lambda}}_0$ and an orthogonal component $\tilde{\boldsymbol{\lambda}}_{1,\perp}$. The null constraint removes the parallel part and retains only the orthogonal derivative component. The overlap parameter $\chi$ quantifies the angle between $\tilde{\boldsymbol{\lambda}}_{0}$ and $\tilde{\boldsymbol{\lambda}}_1$, yielding the retention law $\mathcal I^{\rm null}_{\rm max}/\mathcal I^{\rm lin}_{\rm max}=1-\chi^2$. Note that the figure is schematic and represents response-space geometry rather than the physical TRY layout.}
\label{fig:null_geometry}
\end{figure}

Under the null constraint, the maximum achievable local Fisher information accordingly becomes
\begin{equation}
\mathcal I^{\rm null}_{\rm max}=\boldsymbol{\lambda}^T_1\mathbf{D}^{-1}_0
\boldsymbol{\lambda}_1-\frac{(\boldsymbol{\lambda}^T_0\mathbf{D}^{-1}_0
\boldsymbol{\lambda}_1)^2}{\boldsymbol{\lambda}^T_0\mathbf{D}^{-1}_0
\boldsymbol{\lambda}_0}.
\end{equation}
The second term is the information penalty associated with the background-aligned derivative component. It is exactly the piece removed by the coefficient $\alpha$. This is also the point at which the physical discussion of Sec.~II.C becomes operational: interference visibility and channel structure matter because they control the relative orientation of $\boldsymbol{\lambda}_0$ and $\boldsymbol{\lambda}_1$, and hence the size of the subtraction term.

\subsection{Whitened geometry and exact information-retention law}
A more transparent formulation is obtained by introducing the whitened vectors
\begin{equation}
\tilde{\boldsymbol{\lambda}}_0=\mathbf{D}^{-1/2}_0\boldsymbol{\lambda}_0,\quad \tilde{\boldsymbol{\lambda}}_1=\mathbf{D}^{-1/2}_0\boldsymbol{\lambda}_1.
\end{equation}
These definitions make explicit the inverse-noise geometry already used in Sec.~II.C. In particular,
\begin{equation}
\tilde{\boldsymbol{\lambda}}_0^T\tilde{\boldsymbol{\lambda}}_1=\boldsymbol{\lambda}_0^T\mathbf{D}^{-1}_0\boldsymbol{\lambda}_1,
\end{equation}
and
\begin{equation}
\|\tilde{\boldsymbol{\lambda}}_0\|^2=\boldsymbol{\lambda}_0^T\mathbf{D}^{-1}_0\boldsymbol{\lambda}_0,\quad\|\tilde{\boldsymbol{\lambda}}_1\|^2=\boldsymbol{\lambda}_1^T\mathbf{D}^{-1}_0\boldsymbol{\lambda}_1.
\end{equation}
Thus, the normalized overlap introduced in Sec.~II.C can be written equivalently as
\begin{equation}
\chi=\frac{\tilde{\boldsymbol{\lambda}}_0^T\tilde{\boldsymbol{\lambda}}_1}{\|\tilde{\boldsymbol{\lambda}}_0\|\|\tilde{\boldsymbol{\lambda}}_1\|}.
\end{equation}
The two definitions are in fact identical, since whitening converts the inverse-noise inner product into an ordinary Euclidean inner product.

This representation also clarifies the physical meanings of the two norms. The quantity
\begin{equation}
\|\tilde{\boldsymbol{\lambda}}_1\|^2=\mathcal I_{\rm full}(\theta_0)
\end{equation}
is the full-channel Fisher information, while
\begin{equation}
\|\tilde{\boldsymbol{\lambda}}_0\|^2=\boldsymbol{\lambda}_0^T\mathbf{D}^{-1}_0\boldsymbol{\lambda}_0
\end{equation}
is the noise-weighted magnitude of the nominal background.

Now decompose the derivative response into components parallel and orthogonal to $\tilde{\boldsymbol{\lambda}}_0$:
\begin{equation}
\tilde{\boldsymbol{\lambda}}_1=\tilde{\boldsymbol{\lambda}}_{1,\parallel}+\tilde{\boldsymbol{\lambda}}_{1,\perp},\quad\tilde{\boldsymbol{\lambda}}_{1,\perp}\perp\tilde{\boldsymbol{\lambda}}_0.
\end{equation}
The null constraint removes the parallel component, so that
\begin{equation}
\mathcal I^{\rm null}_{\rm max}=\|\tilde{\boldsymbol{\lambda}}_{1,\perp}\|^2, \quad\mathcal I^{\rm lin}_{\rm max}=\|\tilde{\boldsymbol{\lambda}}_1\|^2.
\end{equation}
Using the orthogonal decomposition,
\begin{equation}
\|\tilde{\boldsymbol{\lambda}}_1\|^2=\|\tilde{\boldsymbol{\lambda}}_{1,\parallel}\|^2+\|\tilde{\boldsymbol{\lambda}}_{1,\perp}\|^2,
\end{equation}
one obtains
\begin{equation}
\frac{\mathcal I^{\rm null}_{\rm max}}{\mathcal I^{\rm lin}_{\rm max}}=1-\chi^2.
\end{equation}
This is the exact information-retention law. It shows that the null constraint removes precisely the fraction $\chi^2$ of the available local Fisher information and retains the complementary fraction $1-\chi^2$. The exact information trade-off implied by the null constraint is summarized in Fig.~2. The retained fraction decreases quadratically as $1-\chi^2$, while the discarded fraction increases as $\chi^2$, showing the cost of null enforcement is controlled entirely by the inverse-noise-metric overlap between the nominal and derivative response vectors.

\begin{figure}[t]
\includegraphics[width=0.95\columnwidth]{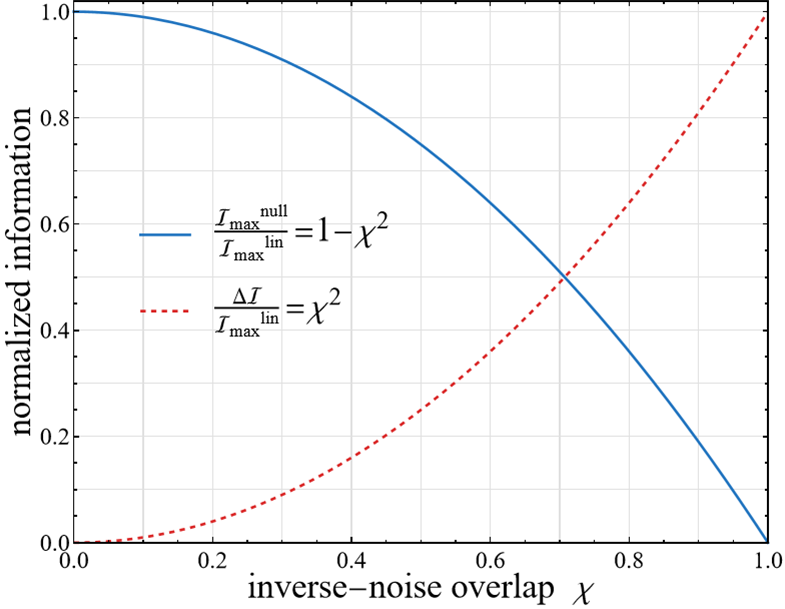}
\caption{Exact information trade-off imposed by the null constraint. 
The retained fraction of local Fisher information is
$\mathcal I^{\mathrm{null}}_{\max}/\mathcal I^{\mathrm{lin}}_{\max}=1-\chi^2$, while the discarded fraction is $\Delta \mathcal I/\mathcal I^{\mathrm{lin}}_{\max}=\chi^2$. Here $\chi$ is the normalized inverse-noise overlap between the nominal response $\boldsymbol{\lambda}_0$ and the derivative response $\boldsymbol{\lambda}_1$. The result is universal; all system dependence enters only through $\chi$.}
\label{fig:null_penalty}
\end{figure}

When $\chi\approx0$, the derivative response is nearly orthogonal to the nominal background, and null enforcement is essentially lossless. When $|\chi|\approx1$, the two are strongly aligned, and most of the information is discarded. Because $\chi$ is the same overlap quantity introduced in Sec.~II.C, its value is shaped by the physical response structure of the TRY channel, including the interference modulation and visibility. In this way, the exact retention law connects the abstract projection geometry directly back to experimentally meaningful channel properties.

\section{Explicit TRY Example and quantitative comparisons}
To illustrate the null-constrained theory in a concrete setting, we return to the phenomenological TRY forward model introduced in Sec.~II.C as well as in Ref.~\cite{Wen2026TRYdiff}. We discretize the source coordinate into $M=121$ equally spaced channels over the interval $[-6,6]$, with spacing $dx=0.1$, and take the following parameter set
\[
N_0\eta=1,\;\;\sigma=1.2,\;\;V=0.85,\;\;\kappa=2.6,\;\;\phi_0=\frac{\pi}{5}.
\]
For simplicity, we choose the operating point $\theta_0=0$. These parameters are not chosen to optimize performance but to provide a smooth and representative channel family in which the structure of the null-coded solution can be displayed transparently.

\subsection{Structure of the optimal null code}
The structure of the optimal null code and its relation to the primitive channel response are demonstrated in Fig.~3. The figure displays the nominal response $\lambda_{0,m}$, the derivative response $\lambda_{1,m}$, the optimal real-valued code $w^{\star}_m$, and its binary approximation. This visualization makes clear how the optimal code inherits the derivative structure while removing the component aligned with the nominal background. The close correspondence is between $w^{\star}_m$ and the inverse-noise-weighted derivative direction $[\mathbf{D}^{-1}\boldsymbol{\lambda}_1]_m$, not $\lambda_{1,m}$ itself. For the present forward model, since $\lambda_{1,m}=[(x_m-\theta_0)/\sigma^2]\lambda_{0,m}$, one has $[\mathbf{D}^{-1}_0\boldsymbol{\lambda}_1]_m=(x_m-\theta_0)/\sigma^2$. Thus, when $\alpha$ is small, the optimal null code is approximately an affine ramp across the source basis.

\begin{figure}[htb]
\includegraphics[width=0.95\columnwidth]{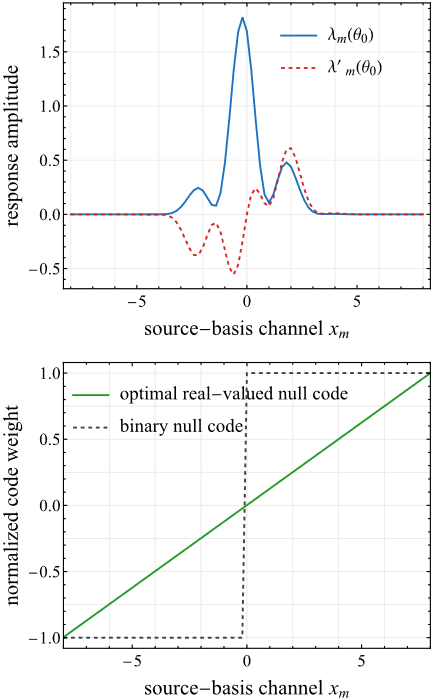}
\caption{Source-basis structure of the null-coded receiver for the explicit TRY model at the operating point $\theta_0=0$. From top to bottom: nominal response $\lambda_{0,m}$, derivative response $\lambda_{1,m}$, optimal real-valued null code $w_m^\star\propto[\mathbf{D}^{-1}_0(\boldsymbol{\lambda}_1-\alpha\boldsymbol{\lambda}_0)]_m$, and its binary approximation $w_m^{(\mathrm{bin})}=\mathrm{sgn}(w^{\star}_m)$. For the present parameter set, the inverse-noise overlap $\chi$ is very small, so the subtraction term $\alpha\boldsymbol{\lambda}_0$ produces only a weak correction to the derivative pattern. Since $\lambda_{1,m}=[(x_m-\theta_0)/\sigma^2]\lambda_{0,m}$, the inverse-noise-weighted derivative direction $\mathbf{D}^{-1}_0\boldsymbol{\lambda}_1$ is approximately ramp-like, and the optimal null code is well captured by by its binary sign structure.}
\label{fig:code_profiles}
\end{figure}

Several features are evident. First, the dominant structure of $w^{\star}_m$ follows that of $\lambda_{1,m}$, confirming that the information content of the measurement is carried by the derivative response. The subtraction term $\alpha\boldsymbol{\lambda}_0$ modifies this pattern only to the extent required to enforce the null condition. Second, for the present parameters, the overlap between $\boldsymbol{\lambda}_0$ and $\boldsymbol{\lambda}_1$ is very weak (as quantified by $\chi$ below), so the projection does not strongly distort the derivative structure. Consequently, the optimal code is already close to a sign pattern. Third, the binary approximation $w^{(bin)}_m$, obtained by retaining only the sign of $w^{\star}_m$, closely tracks the real-valued solution. This indicates that, in regimes where the derivative response is dominated by sign structure rather than amplitude variation, null coding can be implemented using simple two-level source patterns with minimal performance loss. These observations connect directly to the geometric picture of Sec.~IV: the null constraint removes only the background-aligned component, leaving the orthogonal derivative structure largely intact.

\subsection{Information retention and a TRY example}
The performance of the null-coded receiver is quantified by the exact retention law (41). For the representative discretized TRY model considered here with the above parameter set, and symmetric sampling about $\theta_0=0$, the overlap is numerically very small:
\[
\chi\approx-1.23\times10^{-2}.
\]
Accordingly, the null-constrained receiver retains essentially all of the locally available Fisher information,
\[
\frac{\mathcal I^{\rm null}_{\rm max}}{\mathcal I^{\rm lin}_{\max}}\approx0.99985,
\]
so that the discarded fraction is only
\[
\frac{\Delta\mathcal I}{\mathcal I^{\rm lin}_{\max}}=\chi^2\approx1.50\times10^{-4}.
\]
Thus, for this operating point, the null constraint is nearly lossless. On the universal trade-off curve shown in Fig.~2, the TRY example discussed here thus lies extremely close to the $\chi=0$ limit. The small value of $\chi$ reflects the favorable local response geometry of the channel family: the derivative response is already nearly orthogonal to the nominal background in the inverse-noise metric, so only a very weak background-aligned component must be removed by the null projection.

This behavior is also consistent with the code profile in Fig.~3. Because the overlap is so small, the subtraction term $\alpha\boldsymbol{\lambda}_0$ produces only a minor correction to the derivative pattern, and the optimal null code remains visually close to $\boldsymbol{\lambda}_1$.

\subsection{Visibility dependence in the present example}
The detailed dependence of the overlap parameter on the visibility is given in Appendix~A. For the current discretized model, the quantity $\chi(V)$ is determined by the visibility-dependent weighted sums entering
\begin{equation}
\chi(V)=\frac{B_0+VB_1}{\sqrt{(A_0+VA_1)(C_0+VC_1)}}.
\end{equation}
For the symmetric sampling assumed here, $B_0=0$, so
\begin{equation}
\chi(V)=\frac{VB_1}{\sqrt{(A_0+VA_1)(C_0+VC_1)}}.
\end{equation}
This form shows that the overlap vanishes at $V=0$ and, for the given parameter set, remains numerically small throughout the full visibility range $0\leq V\leq1$. As a result, the corresponding retention fact
\[
1-\chi(V)^2
\]
stays very close to unity. In the present illustrative model, visibility reshapes the nominal and derivative response amplitudes and therefore changes the Fisher weights and overlap sums. However, because $\lambda_{1,m}/\lambda_{0,m}=(x_m-\theta_0)/\sigma^2$, the inverse-noise-weighted derivative direction retains a simple ramp-like form, and visibility affects the null penalty primarily through the weighted overlaps rather than a qualitative change of code shape.

This point is important for interpreting the explicit TRY numerics. A high visibility does not by itself imply a large null penalty, nor does a low visibility necessarily imply poor null-constrained performance. What matters is how the visibility-modified channel weights affect the inverse-noise overlap between the nominal and derivative response vectors. In the present geometry, that overlap remains weak, so the null-coded receiver stays close to the nearly lossless regime.

At the same time, the exact expression in Appendix~A makes clear that this behavior is not universal for all parameter choices. Different operating points, phase offsets, fringe frequencies, or asymmetric sampling grids can produce substantially larger values of $|\chi|$, in which case the information-retention curve would depart more visibly from unity. The current example should therefore be interpreted as a favorable local operating regime rather than a generic visibility trend.

\subsection{Robustness under detuning}
The theory developed so far is local in $\theta$, so it is important to examine how a code designed at $\theta_0$ performs away from the point. For a fixed code $\bf w$, we define the local Fisher information at a nearby parameter value as
\begin{equation}
\mathcal I_{\bf w}(\theta)=\frac{(\mathbf{w}^T\boldsymbol{\lambda}_1(\theta))^2}{\mathbf{w}^T\mathbf{D}(\theta)\mathbf{w}}.
\end{equation}
We compare four receiver scenarios:

(i) the full-channel limit $\mathcal I^{\rm lin}_{\rm max}(\theta)$, 

(ii) the optimal null-coded receiver $\mathbf{w}^{\star}$ designed at $\theta_0$, 

(iii) its binary approximation, and 

(iv) representative single-channel monitoring at bright and dark source positions. 

The resulting local Fisher information curves are shown in Fig.~4. For the TRY example considered here, the null-coded receiver is already nearly lossless at the operating point because the overlap parameter $\chi$ is very small. Figure~4 displays that this favorable operating-point geometry also leads to strong robustness under moderate detuning: the optimal null-coded receiver remains close to the full-channel benchmark over a finite neighborhood around $\theta_0$.

\begin{figure}[htb]
\includegraphics[width=1.0\columnwidth]{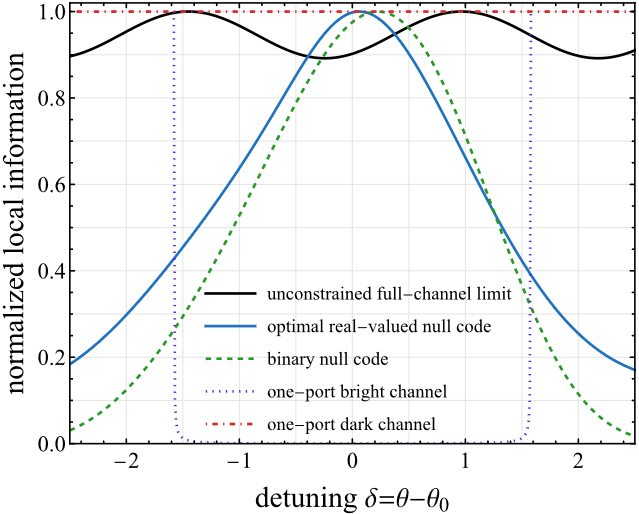}
\caption{Local Fisher information as a function of parameter detuning $\delta=\theta-\theta_0$ for different receiver strategies. Shown are the full-channel limit $\mathcal I^{\rm lin}_{\rm max}(\theta)$, the optimal null-coded receiver $\mathbf{w}^{\star}$, its binary approximation, and representative single-channel (bright and dark) measurements. For the TRY example considered in this work, the null-coded receiver is nearly lossless at the operating point and remains close to the full-channel benchmark over a finite detuning range. The binary approximation follows the optimal null code closely, while the single-channel strategies exhibit substantially reduced local sensitivity.}
\label{fig:detuning_robustness}
\end{figure}

The binary approximation closely tracks the real-valued optimum throughout the same detuning range. This confirms that, in the present region, the dominant information-bearing structure is encoded primarily in the sign pattern of the source-basis code rather than in finite amplitude modulation.

By contrast, the representative bright-channel and dark-channel one-port strategies perform substantially worse. The bright-channel measurement does not reject the nominal background and thus mixes informative and noninformative response components, while the dark-channel measurement remains intrinsically weaker and does not provide the same robust linear sensitivity near the operating point.

These comparisons highlight the practical advantage of source-basis null coding in the TRY architecture: for a favorable local response geometry, it combines nearly full information retention at the operating point with robustness under detuning and compatibility with simple binary implementations.

\section{Binary Codes and Positive-Only Implementation}
The optimal null code derived in Sec.~IV is generally real valued, whereas practical source modulation often imposes discrete or nonnegative constraints. It is thus important to determine how closely the ideal null-coded receiver can be approximated using experimentally accessible source patterns. As already indicated by the explicit example in Sec.~V, the favorable operating-point geometry of the representative TRY channel family leads to an optimal code that is close to a sign pattern, suggesting that high performance should remain accessible even under practical implementation constraints.

\subsection{Binary approximation of the optimal code}
A natural discrete approximation is obtained by retaining only the sign structure of the optimal real-valued code (see Appendix~D). Using the notation of Sec.~IV, this yields the binary pattern
\begin{equation}
w^{(\rm bin)}_m=\mathrm{sgn}\Big([\mathbf{D}^{-1}_0(\boldsymbol{\lambda}_1-\alpha\boldsymbol{\lambda}_0)]_m\Big).
\end{equation}
The corresponding local Fisher information is
\begin{equation}
\mathcal I_{\rm bin}(\theta_0)=\frac{\Big[\big(\mathbf{w}^{(\rm bin)}\big)^T\boldsymbol{\lambda}_1\Big]^2}{\big(\mathbf{w}^{(\rm bin)}\big)^T\mathbf{D}_0\mathbf{w}^{(\rm bin)}}\leq\mathcal I^{\rm null}_{\rm max}(\theta_0),
\end{equation}
where the inequality follows because the real-valued code $\mathbf{w}^{\star}$ is optimal over the full continuous class.

The relevant question is therefore not whether the binary code is exact, but how much performance is lost when only the channel polarity is retained. The example of Sec.~V shows that this loss can be very small: the binary curve in Fig.~4 closely follows the real-valued optimum over the same detuning range, and the sign pattern shown in Fig.~3 already captures the dominant structure of the optimal code. These comparisons imply that the informative content of the receiver is often governed more by the spatial sign structure than by fine amplitude weighting.

This is practically important. It means that the metrological advantage of null-coded sensing is not tied to high-precision analog control of every source channel. In favorable regimes, near-optimal performance can already be obtained using simple two-level source patterns, which substantially lowers the experimental burden.

\subsection{Positive-only realization via sequential coding}
In many experimental implementations, the source intensity must remain nonnegative in each exposure. A signed code must then be realized indirectly by decomposing it into positive and negative parts,
\begin{equation}
w_m=w^{(+)}_m-w^{(-)}_m,
\end{equation}
with
\begin{equation}
w^{(+)}_m=\max(w_m,0),\quad w^{(-)}_m=\max(-w_m,0),
\end{equation}
so that
\begin{equation}
w^{(+)}_m\geq0,\quad w^{(-)}_m\geq0,\quad w^{(+)}_mw^{(-)}_m=0.
\end{equation}
The two nonnegative source patterns $\mathbf{w}^{(+)}$ and $\mathbf{w}^{(-)}$ are then applied sequentially, yielding total detected counts $N_+$ and $N_-$. The signed observable is reconstructed as
\begin{equation}
S_{\rm diff}=N_+-\zeta N_-,
\end{equation}
where $\zeta=1$ is the ideally balanced case, and more generally $\zeta$ can compensate for systematic imbalance between the two exposures, such as drifts in source power, detection efficiency, or acquisition time. A direct positive-only realization of the signed null code in this form is described in Appendix~E.

This construction follows the same basic logic as earlier differential source-basis encoding in the TRY architecture~\cite{Wen2026TRYdiff}, but it is now applied to the optimal null-coded receiver. The null is hence not enforced by requiring any individual exposure to be dark. Rather, each exposure can remain strictly nonnegative and finite-flux, while the null emerges only after differential combination of the two measurements.

The practical significance is that the optimal null-coded receiver remains implementable without negative source intensities and without continuously variable analog modulation. The essential ingredients are only nonnegative source patterns and sequential differencing of the resulting bright measurements.

\subsection{Practical implications}
Taken together, these observations show that source-basis null engineering in the TRY architecture is not simply a formal optimization result. Its core advantages---background cancellation, preserved first-order sensitivity, and strong local information capture---remain accessible even under experimentally realistic modulation constraints.

The binary approximation demonstrates that the optimal receiver is often robust to coarse quantization of the source weights. The positive-only construction shows that even the signed nature of the null-coded observable does not require physically negative modulation; it can be synthesized from ordinary nonnegative exposures. In this sense, the practical receiver preserves the same conceptual structure established in Secs.~III and IV: the useful observable is a coded differential combination of finite-flux source-basis measurements, not a single weak or dark branch.

For this reason, the implementation constraints addressed here do not undermine the main message of the theory. Rather, they strengthen it by showing that the source-basis null can be realized in forms that are experimentally plausible while still retaining the essential performance benefits demonstrated in Sec.~V.

\section{Relation to Existing Null and Superresolution Strategies}

The use of nulls, antisymmetric channels, and derivative-sensitive measurements is a well-established theme in optical metrology and superresolution~\cite{Tsang2016PRX,Nair2016PRL,Paur2016Optica,SanchezSoto2017,Tsang2019PRA,Serabyn1999,Martinod2021NatComm}. Methods such as spatial-mode demultiplexing, parity sorting, and nulling interferometry all build on the same broad insight: near symmetry points, conventional intensity measurements may fail to access the most informative degrees of freedom, whereas suitably engineered observables can isolate the variation that carries local parameter information. In that broad sense, the present work belongs to a larger family of background-rejecting, derivative-sensitive measurement strategies.

At the same time, the specific role played by the null here is different from that in these earlier approaches. The central distinction is not merely that the receiver operates near a dark point, but that the null is engineered directly in a \emph{programmable source basis}. The relevant comparison is therefore not only with nulling strategies in general, but also with where and how the informative measurement basis is physically constructed.

\subsection{Relation to derivative-mode and nulling strategies}
In conventional nulling interferometry and derivative-mode sensing, the measurement basis is typically defined on the detection side, either through optical interference, modal projection, or spatial sorting~\cite{Nair2016OE,SanchezSoto2017,Tsang2019PRA,Serabyn1999,Martinod2021NatComm,Laugier2023}. The receiver isolates informative field components by acting on detected optical modes or image-plane signals after propagation.

By contrast, in the TRY architecture developed here, the measurement basis is defined in the source coordinate through programmable weighting of source channels. The detector remains fixed, and the null is synthesized not by destructive interference at a detection port but by constructing a coded observable whose nominal response cancels while its first-order derivative response is retained. In this sense, detection-basis nulling is a projection performed on optical fields at the receiver, whereas source-basis nulling is a projection performed on the measured channel responses themselves. The two strategies are therefore governed by the same general objective---background rejection with preserved local sensitivity---but they realize that objective at different physical layers of the measurement process.

For this reason, the optimal receiver derived here is best understood as a source-basis realization of derivative-mode sensing~\cite{Nair2016OE,SanchezSoto2017,Tsang2019PRA}. It embodies the same information principle that motivates antisymmetric and derivative-sensitive receivers---namely, isolation of informative variation from a dominant background--but it does so in a different physical layer. The informative observable is engineered through source-basis coding rather than through detection-side mode selection.

This distinction matters both conceptually and practically. Conceptually, it shows that derivative-mode metrology need not be tied to a detector-mode basis. Practically, it opens a route to implementing derivative-sensitive receivers in systems where the source plane is easier to program than the detection optics.

\subsection{Relation to differential and computational imaging}
It is also useful to distinguish the present approach from broader classes of differential and computational measurement strategies. In many computational imaging and differential sensing methods~\cite{Shapiro2008,Bromberg2009,Erkmen2010,Ferri2010,Luo2012,Wen2012,Edgar2019,Sun2019}, useful signals are obtained by subtracting or combining multiple measurements taken under different illumination or acquisition conditions. At a superficial level, the positive-only sequential implementation of Sec.~VI may appear similar, since it also forms an informative observable from a difference of finite-flux measurements.

The key distinction is that the present work is not built around subtraction as a numerical post-processing step alone. Rather, the subtraction is dictated by an explicit metrological optimization problem. The code $\mathbf{w}^{\star}$ is chosen so that the resulting differential observable satisfies a true null condition and maximizes the local Fisher information subject to that constraint. In other words, the differencing rule is not heuristic; it is the optimal solution of a constrained information-theoretic design problem.

This is also what differentiates the present approach from generic computational imaging ideas. The purpose is not to reconstruct a full image or invert a forward operator from multiple coded exposures. Instead, the goal is to engineer a single locally informative observable for parameter estimation. The framework is therefore inherently estimation-theoretic rather than image-reconstruction-based.

\subsection{Relation to structured illumination and resolution enhancement}

Structured illumination and related resolution-enhancement techniques also exploit source-side control~\cite{Gustafsson2000, Gustafsson2005}, so the distinction here should be stated clearly. In those methods, the illumination pattern is designed to mix spatial frequencies and shift otherwise inaccessible spectral content into the passband of the imaging system. Resolution enhancement is then achieved through reconstruction of the redistributed spectral information.

The present work operates on a different principle. The null-coded TRY receiver is not based on frequency mixing, image reconstruction, or recovery of an extended spatial spectrum. Instead, it is formulated entirely as a local estimation problem. The purpose of source modulation here is not to encode high spatial frequencies for later inversion, but to construct an observable that is maximally sensitive to small parameter variations while remaining exactly null at the operating point. Thus, although both approaches involve structured source control, they belong to different conceptual categories: structured illumination modifies spectral accessibility for imaging, whereas source-basis null coding engineers the local information geometry of the measurement itself.

\subsection{Distinct contribution of the present work}
The novelty of the present work can be stated precisely: 

(i) It formulates null-constrained metrology in a programmable source basis within the time-reversed Young architecture. The null is not treated as an incidental dark operating condition, but as a design constraint imposed directly on the coded observable. 

(ii) It derives an explicit and constructive expression for the optimal null code under a general shot-noise-limited channel model. The resulting receiver is the inverse-noise-weighted derivative response with its background-parallel component removed. 

(iii) It establishes an exact and model-independent information-retention law, $\mathcal I^{\rm null}_{\rm max}/\mathcal I^{\rm lin}_{\rm max}=1-\chi^2$, which quantifies the metrological cost of imposing a null in terms of the inverse-noise-metric overlap between nominal and derivative responses. 

(iv) It shows that the resulting strategy remains compatible with experimentally realistic constraints, including binary source patterns and positive-only sequential implementation.

Taken together, these results show that null-coded TRY metrology is not merely another instance of dark-port sensing, nor a reformulation of generic differential or computational imaging ideas. It is a distinct source-basis implementation of derivative-sensitive metrology, made possible by the programmable structure of the TRY platform.

\section{Conclusion}
In summary, we have developed a local theory of null-constrained source-basis metrology in a time-reversed Young interferometer. The central result is that a true metrological null is not simply a dark output, but a coded observable whose nominal response vanishes at the operating point while its first-order sensitivity remains finite.

Within a general shot-noise-limited channel model, the optimal null-constrained receiver is obtained by removing from the inverse-noise-weighted derivative response the component aligned with the nominal background. This yields both a constructive source-basis code and an exact information-retention law, $\mathcal I^{\rm null}_{\rm max}/\mathcal I^{\rm lin}_{\rm max}=1-\chi^2$, which quantifies the metrological cost of imposing the null.

The explicit TRY examples show that, for favorable local response geometry, the null-coded receiver can retain nearly all locally available information while remaining compatible with binary and positive-only implementations. These results establish source-basis null engineering as a distinct and practically meaningful capability of the TRY architecture.

\section*{Acknowledgments}
The author is grateful to Drs.~B. He, Y. Zhai, and D. Zhang, and Ms.~Y. Zhang for encouragement and useful conversations related to the results reported in this work. This work was partially supported by startup funds from Binghamton University and NSF ExpandQISE 2329027.

\appendix

\section{Visibility Dependence of the Overlap Parameter}

In this appendix, we derive an explicit expression for the normalized inverse-noise overlap parameter $\chi$ as a function of the interference visibility $V$ for the discretized TRY channel model introduced in Sec.~II.

Using Eqs.~(8) and (9) of the forward model in Sec. II.C, we evaluate the channel responses at discrete positions $x_m$ and define
\begin{eqnarray}
y_m\equiv x_m-\theta_0, \; g_m\equiv e^{-\frac{y_m^2}{2\sigma^2}}, \; c_m \equiv\cos(\kappa x_m+\phi_0).
\end{eqnarray}
At the operating point $\theta_0$, the nominal and derivative responses [Eqs.~(10) and (11)] can be recast as
\begin{align}
\lambda_{0,m}&=N_0\eta\, g_m(1+V c_m), \nonumber\\
\lambda_{1,m}&=N_0\eta\,\frac{y_m}{\sigma^2}\, g_m(1+V c_m).
\end{align}
Substituting them into the definition of $\chi$ [Eq.~(16)], and using $D_0=\mathrm{diag}(\lambda_{0,m})$, one obtains
\begin{equation}
\chi(V)=\frac{\displaystyle\sum_m\frac{y_m}{\sigma^2}\, g_m(1+Vc_m)}{
\sqrt{\left[\displaystyle\sum_m g_m(1+Vc_m)\right]\left[\displaystyle \sum_m\frac{y_m^2}{\sigma^4}\, g_m(1+Vc_m)\right]}}.
\end{equation}
The overall factor $N_0\eta$ cancels identically.

For notational compactness, one defines
\begin{align*}
A_0&=\sum_mg_m,& A_1&=\sum_mg_mc_m,\\
B_0&=\frac{1}{\sigma^2}\sum_my_mg_m, &B_1&=\frac{1}{\sigma^2}\sum_my_mg_mc_m, \\
C_0&=\frac{1}{\sigma^4}\sum_my_m^2g_m, &C_1&=\frac{1}{\sigma^4}\sum_m y_m^2g_mc_m.
\end{align*}
Then Eq.~(A3) can be written exactly as
\begin{equation}
\chi(V)=\frac{B_0+VB_1}{\sqrt{(A_0+VA_1)(C_0+VC_1)}}.
\end{equation}
In the common case of symmetric sampling about $\theta_0$, one has $B_0=0$, yielding
\begin{equation}
\chi(V)=\frac{VB_1}{\sqrt{(A_0+VA_1)(C_0+VC_1)}}.
\end{equation}
This expression shows that the visibility dependence of $\chi$ is governed by the weighted sums $A_1$, $B_1$, and $C_1$, which depend on the sampling grid, fringe phase $\phi_0$, and spatial frequency $\kappa$. In particular, the overlap is not determined by $V$ alone, but by how the interference modulation reshapes the relative weighting of the source channels in the inverse-noise metric.

Finally, the corresponding information-retention fraction follows directly from
\begin{equation}
\frac{\mathcal I_{\max}^{\mathrm{null}}}{\mathcal I_{\max}^{\mathrm{lin}}}=1-\chi(V)^2,
\end{equation}
which provides an explicit connection between visibility and null-constrained sensitivity.

\section{One-Port Dark Channels and Differential Nulls}
This appendix clarifies why a one-port dark response is not equivalent to the true differential null defined in Sec.~III.

Consider a single nonnegative measurement channel $\mu(\theta)\ge0$ whose mean vanishes at the operating point~$\theta_0$. Expanding around $\theta_0$ with $\delta=\theta-\theta_0$, one typically has
\begin{equation}
\mu(\theta_0+\delta)=a\delta^2+O(\delta^4),\quad a>0.
\end{equation}
The derivative of the mean is then
\begin{equation}
\frac{d\mu(\theta_0+\delta)}{d\theta}=2a\delta+O(\delta^3).
\end{equation}
Under a Poisson model, the corresponding Fisher information is
\begin{equation}
\mathcal{I}_{\mathrm{1ch}}(\theta)=\frac{[d\mu(\theta)/d\theta]^2}{\mu(\theta)}.
\end{equation}
Substituting the local expansions (B1) and (B2) gives
\begin{equation}
\mathcal{I}_{\mathrm{1ch}}(\theta_0+\delta)=\frac{(2a\delta)^2}{a\delta^2}+O(\delta^2)=4a+O(\delta^2),
\end{equation}
for $\delta\neq 0$. Thus, a dark point in a single channel does not by itself imply vanishing Fisher information. However, this does not make it equivalent to a useful differential null. The essential problem is that the leading response is even:
\begin{equation}
\mu(\theta_0+\delta)=\mu(\theta_0-\delta).
\end{equation}
As a result, the measurement does not distinguish the sign of the parameter shift at leading order and therefore cannot serve as a signed linear local readout. In addition, as $\delta\to 0$, the useful signal is concentrated in a weak branch, making the operating point more fragile with respect to background contamination, detector imperfections, and model mismatch.

By contrast, the differential null used in Sec.~III satisfies
\[
\mathbf{w}^T\boldsymbol{\lambda}_0=0,\quad\mathbf{w}^T\boldsymbol{\lambda}_1\neq0,
\]
so that
\begin{equation*}
\langle S_{\mathbf{w}}(\theta_0+\delta)\rangle=\mathbf{w}^T\boldsymbol{\lambda}_1\,\delta+O(\delta^2).
\end{equation*}
The leading response is therefore linear and signed in $\delta$. The null arises not from a vanishing one-port intensity, but from a differential combination of finite-flux channels whose nominal contributions cancel while the derivative-bearing part survives. This is the sense in which a true metrological null is fundamentally different from one-port darkness.

\section{Derivation of the Optimal Null Code}
\label{app:optimal_null_code}

In this appendix we derive the source-basis code that maximizes the local sensitivity under the null constraint. The problem is to optimize the quadratic ratio
\begin{equation}
\mathcal I(\mathbf{w})=\frac{\left(\mathbf{w}^{T}\bm{\lambda}_{1}\right)^{2}}
{\mathbf{w}^{T}\mathbf{D}_{0}\mathbf{w}},
\end{equation}
subject to $\mathbf{w}^{T}\bm{\lambda}_{0}=0$ (Eq.~(20)). Here, again, $\bm{\lambda}_{0}=\bm{\lambda}(\theta_{0})$ is the nominal response vector at the operating point, $\bm{\lambda}_{1}=\partial_{\theta}\bm{\lambda}(\theta)|_{\theta=\theta_{0}}$ is the corresponding parameter-derivative vector, and $\mathbf{D}_{0}=\mathrm{diag}\!\left(\lambda_{0,1},\lambda_{0,2},\dots,\lambda_{0,M}\right)$
is the diagonal shot-noise matrix at $\theta_{0}$.

Because the functional in Eq.~(C1) is homogeneous in $\mathbf{w}$, the overall scale of the code does not affect the optimization. It is therefore convenient to impose the normalization
\begin{equation}
\mathbf{w}^{T}\mathbf{D}_{0}\mathbf{w}=1,
\end{equation}
in which case maximizing $\mathcal I(\mathbf{w})$ is equivalent to maximizing the linear quantity $\mathbf{w}^{T}\bm{\lambda}_{1}$ under the two constraints in Eqs.~(20) and (C2). 
We therefore introduce the Lagrangian
\begin{equation}
\mathcal{L}=\mathbf{w}^{T}\bm{\lambda}_{1}-\frac{\beta}{2}\left(\mathbf{w}^{T}\mathbf{D}_{0}\mathbf{w}-1\right)-\gamma\,\mathbf{w}^{T}\bm{\lambda}_{0},
\end{equation}
where $\beta$ and $\gamma$ are Lagrange multipliers.

Stationarity with respect to $\mathbf{w}$ gives
\begin{equation}
\frac{\partial\mathcal{L}}{\partial\mathbf{w}}=\bm{\lambda}_{1}-\beta\mathbf{D}_{0}\mathbf{w}
-\gamma\bm{\lambda}_{0}=0.
\end{equation}
Solving for $\mathbf{w}$ yields
\begin{equation}
\mathbf{w}=\beta^{-1}\mathbf{D}_{0}^{-1}\left(\bm{\lambda}_{1}-\gamma\bm{\lambda}_{0}\right).
\end{equation}

The multiplier $\gamma$ is then fixed by the null condition. Substituting Eq.~(C5) into Eq.~(20) gives
\begin{equation}
\bm{\lambda}_{0}^{T}\mathbf{D}_{0}^{-1}\left(\bm{\lambda}_{1}-\gamma\bm{\lambda}_{0}\right)=0,
\end{equation}
from which
\begin{equation}
\gamma=\frac{\bm{\lambda}_{0}^{T}\mathbf{D}_{0}^{-1}\bm{\lambda}_{1}}{
\bm{\lambda}_{0}^{T}\mathbf{D}_{0}^{-1}\bm{\lambda}_{0}}.
\end{equation}
Identifying this coefficient with the quantity denoted by $\alpha$ in the main text, we obtain the optimal null-code direction
\begin{equation*}
\mathbf{w}^{\star}\propto\mathbf{D}_{0}^{-1}\left(\bm{\lambda}_{1}-\alpha\bm{\lambda}_{0}\right),
\quad\alpha=\frac{\bm{\lambda}_{0}^{T}\mathbf{D}_{0}^{-1}\bm{\lambda}_{1}}{
\bm{\lambda}_{0}^{T}\mathbf{D}_{0}^{-1}\bm{\lambda}_{0}},
\end{equation*}
which are the exact expressions (27) and (28) quoted in Sec.~IV.B of the main text.

This result has a simple interpretation. The factor $\mathbf{D}_{0}^{-1}$ weights each source-basis channel by the inverse of its shot-noise variance, while the subtraction of the $\alpha\bm{\lambda}_{0}$ term removes the component of the derivative response that lies parallel to the nominal background in the inverse-noise metric. 
The optimal null code therefore retains only the derivative component that survives the null constraint.

Finally, substituting the above result back into the Rayleigh quotient in Eq.~(C1) gives the maximal null-constrained sensitivity
\begin{equation*}
\mathcal I_{\mathrm{null}}^{\max}=\bm{\lambda}_{1}^{T}\mathbf{D}_{0}^{-1}\bm{\lambda}_{1}
-\frac{\left(\bm{\lambda}_{0}^{T}\mathbf{D}_{0}^{-1}\bm{\lambda}_{1}\right)^{2}}{
\bm{\lambda}_{0}^{T}\mathbf{D}_{0}^{-1}\bm{\lambda}_{0}},
\end{equation*}
which is the result (31) given in the main text to quantify the information retained under the null constraint.

\section{Sign of Binary Approximation}
The binary approximation to the optimal null code is obtained by taking the sign of the optimal real-valued solution:
\begin{equation}
w_m^{(\mathrm{bin})}=
\begin{cases}
+1,&\left[\mathbf{D}_0^{-1}(\boldsymbol{\lambda}_1-\alpha\boldsymbol{\lambda}_0)\right]_m>0, \\
-1,&\left[\mathbf{D}_0^{-1}(\boldsymbol{\lambda}_1-\alpha\boldsymbol{\lambda}_0)\right]_m<0, \\
0,&\left[\mathbf{D}_0^{-1}(\boldsymbol{\lambda}_1-\alpha\boldsymbol{\lambda}_0)\right]_m=0.
\end{cases}
\end{equation}
This expression makes explicit that the binary code preserves only the sign structure of the optimal null code. As discussed in Sec.~VI, this approximation is effective when the optimal solution is close to a sign pattern, which typically occurs when the overlap $\chi$ is small.

\section{Positive-Only Realization of a Signed Null Code}
The optimal null code derived in Sec.~IV is generally signed. In many source-modulation platforms, however, each exposure must remain nonnegative. This appendix shows explicitly how the null-coded observable can nevertheless be implemented using only positive source patterns.

For any real code vector $\bf w$, one defines its positive and negative parts by $w^{(+)}_m$ and $w^{(-)}_m$. Equivalently,
\begin{equation}
\mathbf{w}=\mathbf{w}^{(+)}-\mathbf{w}^{(-)}.
\end{equation}
The two nonnegative patterns $\bf w^{(+)}$ and $\bf w^{(-)}$ can then be applied in two sequential measurements. Let the corresponding detected counts be $N_+$ and $N_-$. A signed coded observable is reconstructed as $S_{\rm diff}=N_+-\zeta N_-$. The expectation value then becomes
\begin{equation}
\langle S_{\rm diff}(\theta)\rangle=\big(\mathbf{w}^{(+)}-\zeta\mathbf{w}^{(-)}\big)^T\boldsymbol{\lambda}(\theta).
\end{equation}
In the balanced ideal case $\zeta=1$, this reduces to
\begin{equation}
\langle S_{\rm diff}(\theta)\rangle=\mathbf{w}^T\boldsymbol{\lambda}(\theta),
\end{equation}
so the sequential implementation reproduces exactly the same mean response as the signed code. In particular, if $\mathbf{w}=\mathbf{w}^{\star}$, then the differential observable inherits the null condition and first-order sensitivity as established in Secs.~III and IV.

This construction makes clear that the null does not require any single exposure to be dark, nor does it require physically negative source intensity. Each exposure remains nonnegative and finite-flux; the null emerges only after differential recombination. This is fully consistent with the source-basis interpretation of the TRY architecture and with the earlier differential source-basis framework~\cite{Wen2026TRYdiff} on which the present implementation builds.

A practical consequence is that the signed structure of the optimal null-coded observable can be realized using standard nonnegative modulation hardware, with the main experimental requirements being sequential stability and calibration of the balancing factor $\zeta$. In this sense, the positive-only construction is not an approximation to the theory, but a direct physical implementation of the same coded measurement principle.

\end{document}